\documentclass[pdftex,aps,amssymb,prl, showpacs,twocolumn]{revtex4}
\usepackage{amsmath}
\usepackage{amssymb}
\usepackage{amsthm}
\usepackage[pdftex]{graphicx}
\usepackage{pslatex}
\usepackage{booktabs}
\usepackage{fancyhdr}

\begin{document}
\title{Measuring the true quality factor of an ultrafast photonic microcavity: homogeneous versus inhomogeneous broadening}

\author{Alex Hartsuiker}\email{hartsuiker@amolf.nl}
\affiliation{Center for Nanophotonics, FOM Institute for Atomic
and Molecular Physics (AMOLF), Kruislaan 407, 1098 SJ
Amsterdam,
The Netherlands}%
\author{Allard P. Mosk}
\affiliation{Complex Photonic Systems (COPS), MESA+ Institute
for Nanotechnology, University of Twente, The Netherlands}
\author{Julien Claudon}
\author{Jean-Michel G\'erard}
\affiliation{CEA-CNRS, "Nanophysics and Semiconductors" joint
laboratory, CEA/INAC/SP2M, 17 rue des Martyrs, 38054 Grenoble
Cedex 9, France}
\author{Willem L. Vos}
\affiliation{Center for Nanophotonics, FOM Institute for Atomic
and Molecular Physics (AMOLF), Kruislaan 407, 1098 SJ
Amsterdam,
The Netherlands}%
\affiliation{Complex Photonic Systems (COPS), MESA+ Institute
for Nanotechnology, University of Twente, The Netherlands}

\date{\today, version 1}
\begin{abstract}
We have measured time-resolved the photon storage time and the
quality factor of an ultrafast photonic cavity using an
autocorrelator. The cavity consists of a $\lambda$-thick GaAs
layer sandwiched between GaAs/AlAs distributed Bragg reflectors
and resonates at 985 nm wavelength. The inverse relative
linewidth measured with white light reflectivity is 830, while
the quality factor obtained from the time resolved measurements
is 1500. The photon storage time in the cavity is 0.78 ps. We
show that the difference between the quality factor and the
inverse relative linewidth results from inhomogeneous
broadening of the microcavity resonance due to a spatial
gradient in the cavity layer.
\end{abstract}
\maketitle


\section{Introduction}


Since light is extremely elusive there is a great interest to
store photons in a small volume for a certain time. Storage of
photons in an applicable way can be achieved using solid state
cavities. Tanabe \emph{et al.} used cavities to create large
pulse delays with small group velocities by storing light in a
cavity inside a 2D photonic crystal slab
\cite{Tanabe2007ab,Tanabe2007ac}. Another application where
storage of light in a cavity plays a crucial role is changing
the color of light as was studied by Preble \emph{et al.}
\cite{Preble2007aa}. Ultimately, with a microcavity the strong
coupling regime of cavity quantum electro dynamics can be
entered \cite{Reithmaier2004aa,Yoshie2004aa}. In the strong
coupling regime a cavity and a two level system together form a
new set of states. Normal-mode splitting of a coupled
exciton-photon mode was observed in a planar microcavity
\cite{Weisbuch1992aa}. Other interesting experiments have been
performed on planar cavities, e.g Bose-Einstein condensation of
exciton polaritons \cite{Kasprzak2006aa} and the investigation
of the limitations of a scanning Fabry-P\'{e}rot interferometer
\cite{Marzenell2000aa}.




An important characteristic parameter of a cavity resonance is
the storage time of light $\tau_{cav}$. The storage time is
defined by the response of the cavity resonance to a Dirac
pulse. Excitation of the electromagnetic field in a cavity was
 studied in \cite{Erden2008aa}. The
response to the Dirac pulse is given by an exponential decay of
the intensity \emph{I(t)} in the cavity resonance
\cite{Feynman1964aa}:
\begin{equation}
I(t)=I_0 e^{-t/\tau_{cav}}, \label{eq:cavdecay}
\end{equation}
with $I_0$
 the initial intensity that the
 pulse stores in the cavity. However, in more complex cavities the behavior of the cavity can be very different
 from the single exponential case \cite{Hart2009aa}. To compare cavities
independent of their resonance frequencies $\omega_0$, the
widely used figure of merit is the resonance quality factor
\emph{Q}, which is defined as:
\begin{equation}
Q \equiv \tau_{cav}\omega_{0}. \label{eq:defQ}
\end{equation}
 Physically, the quality factor is proportional to the ratio between the total energy stored and the energy lost per
 cycle. At optical frequencies a cavity with a feasible high quality factor of $Q = 10^6$ is relatively slow
 with a response time in the order of nanoseconds. A
 cavity with a moderate quality factor $Q = 1000$, however, is fast with a
 response time of picoseconds. The picosecond timescale allows
 ultrafast access and storage of light in the cavities.

A common procedure to estimate the quality factor of a cavity
is to measure a transmission or reflectivity spectrum and
extract \emph{Q} from the relative linewidth of the cavity
resonance
\cite{Gerard1998aa,Reithmaier2004aa,Yoshie2004aa,Yoshie2001aa,Asano2006aa,Preble2007aa,Harding2007aa}.
For a single resonance without dephasing, one can use the
Wiener-Khintchine theorem, which relates the field
autocorrelate to the intensity spectrum, to obtain
\begin{equation}
Q=\frac{\omega_0}{\Delta\omega}. \label{eq:QisinvLinewidth}
\end{equation}
However, if there is significant dephasing, e.g. due to
inhomogeneous broadening or thermal noise, $\Delta\omega$ will
in general be larger and $Q>\omega_0/\Delta\omega$.

From many resonating systems in condensed matter and solid
state physics, it is known that besides homogeneous broadening
there is also the possibility of inhomogeneous broadening of a
resonance \cite{Lagendijk1993aa,Demtroder1996aa}. In the case
of an ensemble of resonators inhomogeneous broadening of a
resonance results from inhomogeneities in the resonance
frequency. If the resonance frequency is different for each
resonator the linewidth of the ensemble is broader than the
linewidth of a single resonator and the ensemble linewidth is
typically determined by the distribution of resonance
frequencies. In the case of inhomogeneous broadening the
linewidth will only give a lower boundary for the range of
possible \emph{Q} values. The true quality factor must in this
case be determined from dynamic measurements.

 A dynamic measurement to determine the quality factor is a cavity
 ring down experiment as was treated in \cite{Armani2003aa}. In this case a
 cavity is excited by a pulse and the intensity emitted from the cavity is
 measured as a function of time. In the case of storage times in the order of nanoseconds and
very high quality factors ($Q =10^6$) time correlated single
photon counting can be used to determine the storage time
\cite{Tanabe2007ab}. In our case of ultrafast cavities that
decay on a ps timescale with moderate quality factor
($Q=1000$), an intensity autocorrelation function is the method
of choice for determining the quality factor.

The normalized correlation function that is measured is an
intensity autocorrelation function $G^2$, which is given by
\cite{Diels1996aa}
\begin{equation}
G^2(\tau)=\frac{<I(t)I(t-\tau)>}{I_0^2},
\label{eq:2ndOrderAutocorrelation}
\end{equation}
where $\tau$ is the delay time between the pulses from each of
the interferometer branches, $I_0^2$ is equal to maximum value
of the unnormalized autocorrelation value, and $I(t)$ is the
time dependent intensity. There is no phase in equation
\ref{eq:2ndOrderAutocorrelation}, which means that this is the
proper autocorrelation function, also in case of dephasing. The
autocorrelate has its maximum at delay $\tau =0$, when the
pulses in the two branches of the Michelson interferometer
overlap. For example the autocorrelate of a Gaussian pulse is
given by a Gaussian shape, where the width of the input pulse
$\tau_{ip}$ and the autocorrelate are related as $\tau_{ac} =
\sqrt(2)\tau_{ip}$. From the autocorrelate of a pulse stored in
the cavity resonance, the storage time can be found from the
full width at half maximum $\tau_{FWHM}$ of $G^2$, with
$\tau_{cav} = 0.63\tau_{FWHM}$.



\section{Experimental}
Our structure is a planar cavity that consists of a GaAs
$\lambda$-thick layer (277 nm thick), sandwiched between two
Bragg stacks. One Bragg stacks consists of 12 and the other
Bragg stack consists of 16 pairs of $\lambda/4$-thick layers of
nominally pure GaAs or AlAs. The same structure was studied in
Ref. \cite{Harding2007aa}. The sample was grown at CEA in
Grenoble by means of molecular beam epitaxy at 550$^{o}$C
\cite{Gerard1996aa}. For experiments outside the present scope
the sample was doped with $10^{10}$cm$^{-2}$ InGaAs/GaAs
quantum dots, which hardly influence our experiment
\footnote{The maximum unbroadened refractive index change of
the dots amounts to only $10^{-8}$, while the absorption at
resonance is less than $0.02$ cm$^{-1}$.}. There is a spatial
gradient in the cavity thickness of $\frac{\delta d}{\delta x}
= 5.64$ nm/mm \cite{Harding2008aa}. The spatial gradient
results in a position dependent resonance frequency. In our
measurements we average the transmitted intensity over the area
of the focal spot. The different resonance frequencies cause
the resonance to broaden inhomogeneously.

White-light reflectivity and transmission were measured with a
broadband white-light spectrometer setup with a spectral
resolution of about $\Delta\lambda = 0.2$ nm
\cite{Thijssen1999aa}. The transmission spectrum was measured
with a collimated beam with a diameter of 2 mm. The
reflectivity spectrum was measured with a glass objective with
a numerical aperture NA = 0.05 and a focus diameter of $100$
$\mu m$.
 The reflectance spectrum of a gold
mirror was used as a reference.

For pulse transmission and the intensity autocorrelate, we used
a Titanium Sapphire laser that emits $\tau_{ip}=0.115$ ps
pulses at $\lambda = 800 $ nm at a repetition rate of 1 kHz
(Hurricane, Spectra Physics). The laser drives an optical
parametric amplifier (OPA, Topas 800-fs, Light Conversion),
which generates the pulses used to probe the photonic cavity.
The center wavelength of the OPA pulses can be tuned between
450 nm and 2400 nm. We used a fiber optic spectrometer
(USB2000, Ocean Optics) to measure transmission spectra of the
femtosecond pulses. We measured with an unfocused collimated
beam with a spot diameter of 2 mm, and a numerical aperture NA
= $10^{-4}$. The intensity autocorrelation function was
measured using a Pulse Check autocorrelator (APE GmbH). The
autocorrelator consists of a Michelson interferometer with a
scanned delay path and a nonlinear crystal that generates
second harmonic light. The autocorrelator has a maximum range
of 15 ps with a resolution of 1 fs. We used the same beam
parameters as in transmission. The intensity on the sample is
100 $kWcm^{-2}$, sufficiently low to avoid non-linear effects.

Simulations were performed with the finite-difference
time-domain (FDTD) method using a freely available software
package with subpixel smoothing for increased accuracy
\cite{Farjadpour2006aa}.


\section{Experimental results}


\begin{figure}[htb]
  \begin{center}
  \includegraphics[width=8cm]{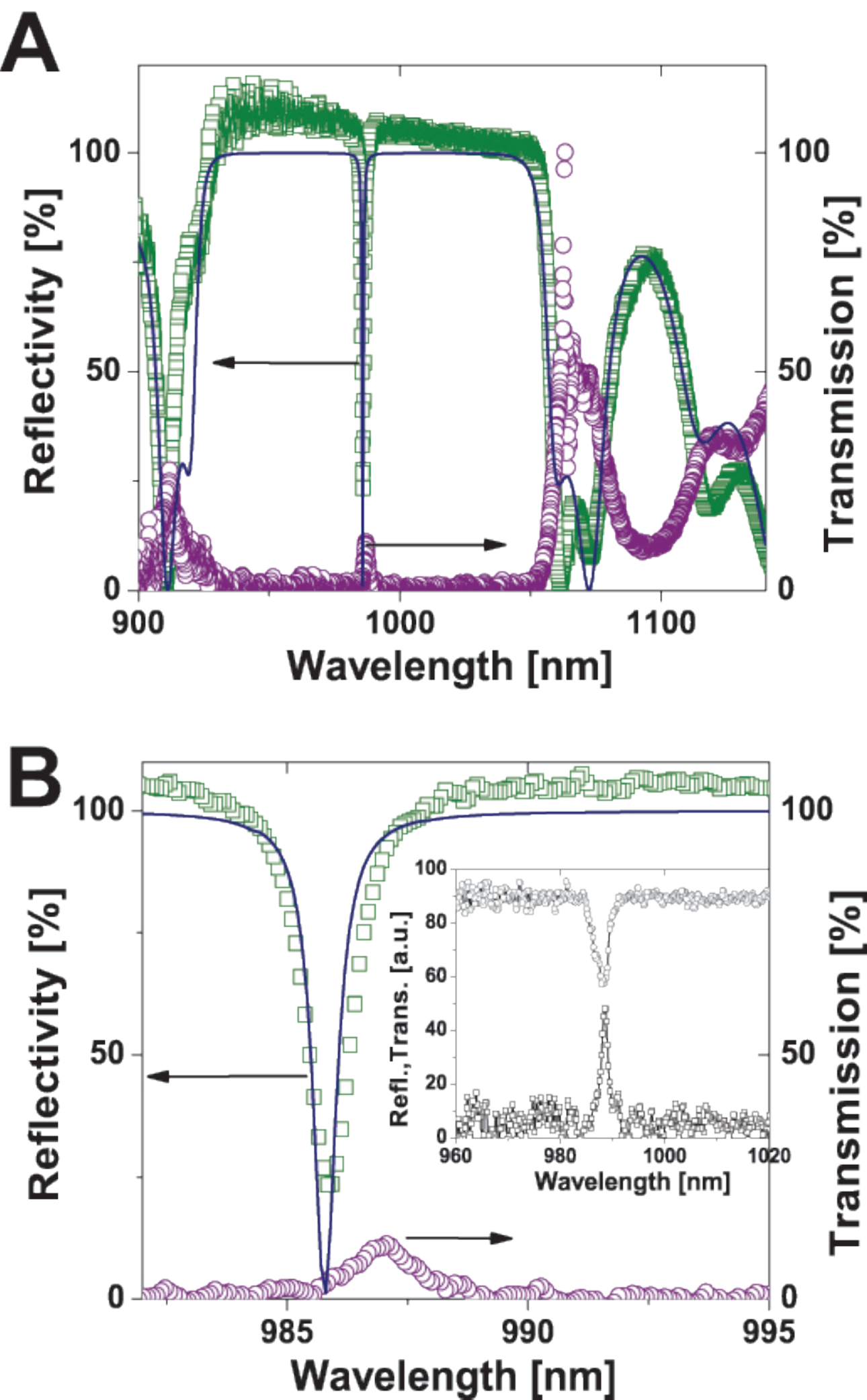}
  \caption{\emph{(A) Linear reflectivity and transmission spectrum of the GaAs/AlAs microcavity. The solid line
  represents the fit with a Transfer Matrix (TM) model. A stop band is apparent in both reflection and
  transmission; the trough in the reflectivity spectrum
  and the peak in the transmission spectrum reveal the presence of the cavity. (B) Zoom-in of (A). From the linewidth of the
  trough and peak, an inverse relative linewidth of 830 was found in reflection and transmission. The resonance
  is slightly shifted between the transmission and reflection measurement due to realignment of the sample
  between the measurements. The inset in B shows a reflection and a transmission spectrum, measured at the
  same position. Here the trough and peak
  are clearly at the same wavelength.   }}\label{071123ReflTrans}
  \end{center}
\end{figure}

In figure \ref{071123ReflTrans} A we show the reflection and
transmission spectra of the planar cavity. A prominent stopband
with a reflection of 100 \% and a transmission of 0 \% is
visible. Outside the stopband a Fabry-P\'{e}rot fringe pattern
is visible, while inside the stopband a narrow trough in
reflection and a narrow peak in transmission mark the position
of the cavity resonance. An effect of the spatial gradient in
the cavity thickness is visible in the spectra in figure
\ref{071123ReflTrans} B: The frequencies of the peak and
trough, which are measured at different sample position, differ
slightly. Reflectivity and transmission measurements on the
same spot are shown as an inset in figure \ref{071123ReflTrans}
B. The trough and the peak are clearly at the same wavelength
as expected.

The solid line in figures \ref{071123ReflTrans} A and
\ref{071123ReflTrans} B represents a transfer matrix (TM)
calculation, with fixed complex input parameters $n_{GaAs}$
\cite{Blakemore1982aa} and $n_{AlAs}$ \cite{Fern1971aa}. The
thickness of the $\lambda/4$ layers ($d_{GaAs} = 70.2 $ nm and
$d_{AlAs} = 83.2 $ nm) and the thickness of the cavity
($d_{cav} = 277 $ nm) were obtained by fitting the results of
the calculations to the measured spectrum. These values are in
agreement with expected values from the fabrication process.
The calculation fits well with respect to frequency and
amplitude. The reflectivity of the measured stopband is higher
than the calculated value of 100 \% because of a small
systematic error in the gold reference spectrum.


It is apparent from figure \ref{071123ReflTrans} B that the
calculated linewidth of the cavity resonance is narrower than
the measured linewidth. We attribute this discrepancy to
inhomogeneous broadening of the measured linewidth, due to the
spatial gradient in the cavity layer thickness. With the 100
$\mu m$ diameter spot we average over different positions and
therefore over different resonance frequencies. Broadening due
to a spread in wavevectors can be neglected since the numerical
aperture of the impinging beam was made very small ($NA<0.05$),
as opposed to \cite{Harding2007aa}, where a high NA was used.
We find that the relative linewidth in both reflection and
transmission equals $ \frac{\lambda_0}{\Delta\lambda}$ = 830,
with $\Delta\lambda$ the full width at half maximum (FWHM) and
$\lambda_0$ the resonance wavelength. The transfer matrix
calculation yields an inverse relative linewidth of $1640
\pm100$, about double the value of the inverse linewidth
measured with white-light spectroscopy.

\begin{figure}[htb]
  \begin{center}
  \includegraphics[width=8cm]{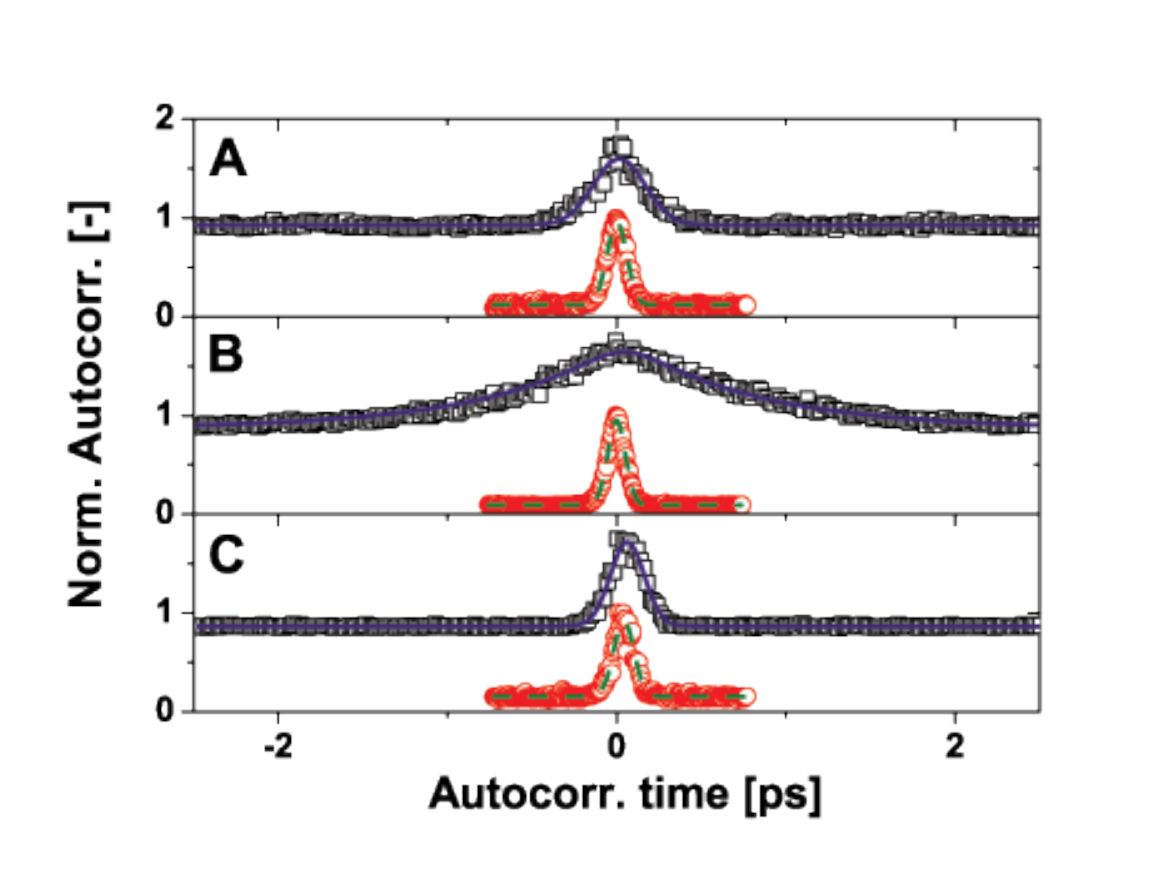}
  \caption{\emph{Normalized intensity autocorrelation traces of pulses transmitted through a planar cavity
   at different OPA wavelength settings:
  930 nm (A), 985 nm (B) and 1070 nm (C). The autocorrelation traces of the input pulses are given by the
  circles, while the autocorrelation traces of pulses transmitted through the cavity
  are offset by 0.9 and given by squares. The dashed and solid lines are fits to the autocorrelation traces,
  without and with sample respectively.
   The shape of the autocorrelation trace is Gaussian for the pulses
 from the OPA. The pulses that are on resonance with the cavity show
  an autocorrelate that agrees very well with the autocorrelation trace from the damped oscillator model (B).
  The shape of the pulses transmitted through a
  non-photonic range of the sample
  remains Gaussian.} }\label{071123WidthACF5psPanel}
  \end{center}
\end{figure}

We measured the intensity autocorrelation traces to determine
the true storage time and \emph{Q} of the cavity resonance with
a time-resolved measurement. Figure
\ref{071123WidthACF5psPanel} shows the autocorrelation traces
at values of the center wavelength of the OPA, $\lambda_{OPA} =
930 $ nm (A), $\lambda_{OPA} = 985 $ nm (B) and $\lambda_{OPA}
= 1070 $ nm (C).
All figures show that the pulses that are transmitted through
the sample are broader than the input pulses. The width of the
input pulses is $\tau_{ip}=0.115$ ps and the shape Gaussian,
which we expect from the specifications of our laser system.
The transmitted pulses are broadened by dispersion in the off
resonance cases (A) and (C). In the case of figure
\ref{071123WidthACF5psPanel} B the broadening is the result of
the storage of the photons in the cavity.

The shape of the autocorrelation trace of the transmitted
pulses is Gaussian for pulses transmitted outside the stop
band, as expected. The autocorrelation traces measured on
resonance with the cavity (B) are non-Gaussian. This is typical
for autocorrelation traces near the cavity resonance, because
of the exponential decay of the energy stored in the cavity.
The autocorrelation traces calculated with a damped oscillator
model is shown in figure (B) and fits the experimental data
very well. From the width of the autocorrelation trace on
resonance ($\tau_{FWHM}=1.1$ ps), we conclude that the true
storage time of our cavity is $\tau_{cav} = 0.78\pm0.05$ ps.



\begin{figure}[htb]
  \begin{center}
  \includegraphics[width=8cm]{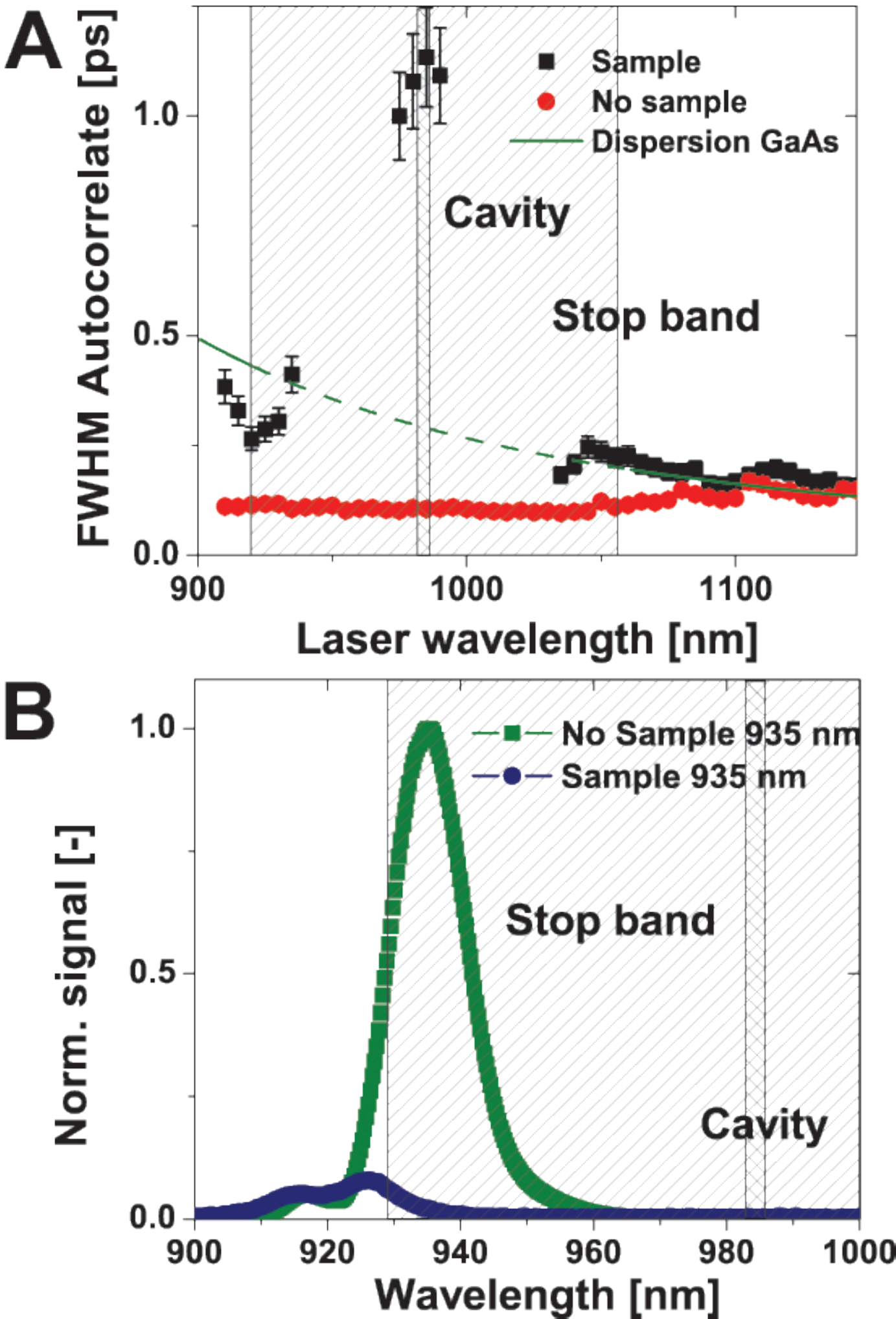}
  \caption{\emph{(A) Pulse width as a function of wavelength setting. The pulses that passed through the sample (squares) are broadened
  with respect to the input pulses (circles). The solid/dashed line represents the FWHM of a 15 nm spectrally wide
  input pulse that is transmitted through a 350 $\mu m$ GaAs wafer that is
  used as a substrate. (B)
  Normalized spectrum transmitted through the microcavity, and reference spectra of
  pulses directly from the OPA. For 935 nm wavelength setting (green squares) the transmitted spectrum
  consists of the tail of the input spectrum that continues in the region on the blue side of the stop band (blue circles).   } }\label{071123WidthACFvsWL}
  \end{center}
\end{figure}

 To further analyze the autocorrelation traces we plot the full width at half maximum
 of the measured autocorrelation traces. The results are shown in figure
 \ref{071123WidthACFvsWL} A
 as a function of center wavelength of the laser.
The width of the autocorrelate of pulses without the sample is
$\tau_{ac}= 0.115\times\sqrt{2}$ ps and essentially independent
of laser wavelength, as expected from the OPA specifications.
 In the presence of the sample, we observe a more complex dependency on the wavelength, with
 three regimes: Transmission on resonance, transmission outside the stopband, transmission inside the stopband.
  Near the cavity resonance the width of the autocorrelate increases drastically to
 $\tau_{FWHM}=1.1$ ps. The width of the autocorrelate at the cavity resonance is
attributed to the storage of light in the cavity: The storage
time of the cavity $\tau_{cav} = 0.78\pm0.05$ ps and the
quality factor is equal to $1500\pm100$.

Outside the stopband the pulses are broadened. The width of the
autocorrelate outside the stop band is about
$0.2\times\sqrt{2}$ ps. We attribute the broadening outside the
stopband region to dispersion in the GaAs substrate. From
figure \ref{071123WidthACFvsWL}A it can be seen that the width
of the autocorrelation traces matches well the expected width
for a pulse transmitted through a GaAs wafer
\cite{Saleh1991aa}. The expected width is calculated for a GaAs
wafer with a thickness of 350 $\mu m$, from the dispersion
given by Blakemore \cite{Blakemore1982aa}.

In figure \ref{071123WidthACFvsWL} A, we observe datapoints
inside the stopband, where a transmission of 0 \% is expected.
We measure values for the width that are close to the values
outside the stopband. The situation in this case is sketched in
figure \ref{071123WidthACFvsWL} B where we see the transmitted
spectrum with and without sample. We observe that the blue part
of the spectrum is transmitted, which means that the measured
with of the intensity autocorrelate is the value for the blue
side of the stopband.

\section{Modeling}

To obtain a physical picture of the decay mechanism inside the
cavity and to verify what the true \emph{Q} is, we model the
behavior with a damped harmonic oscillator. We furthermore
performed FDTD calculations to calculate the \emph{Q} of the
cavity in the ideal case and to  check the validity of the
harmonic oscillator model \footnote{We used $e^{-t\omega_0/Q}$
as the impulse response of intensity in the cavity.}.

\begin{figure}[htb]
  \begin{center}
  \includegraphics[width=8cm]{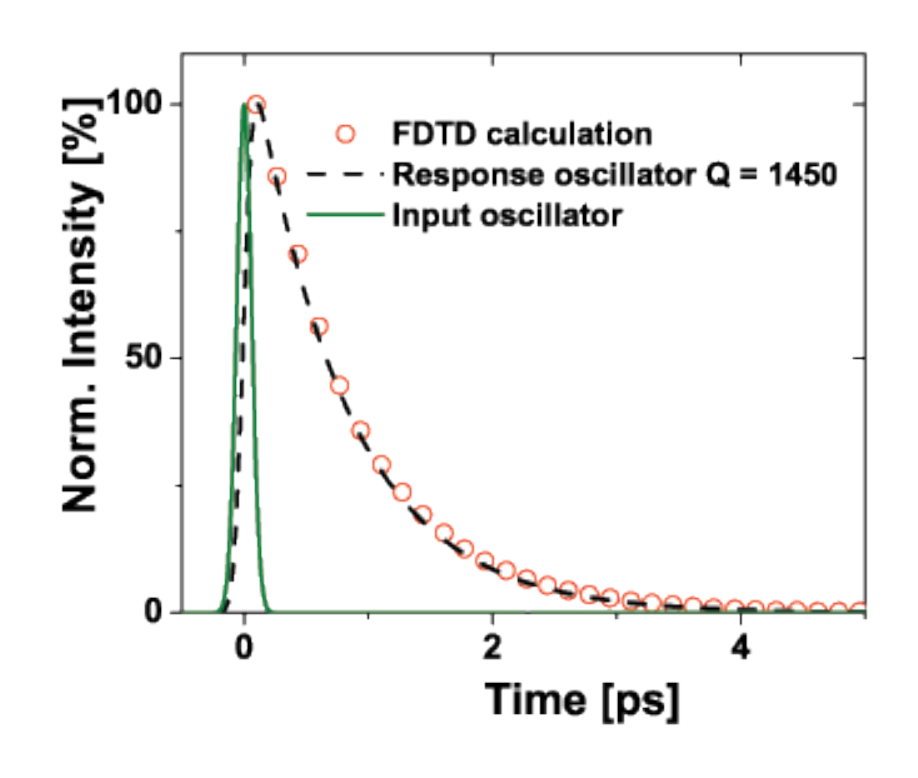}
  \caption{\emph{Response of a damped harmonic oscillator with a quality factor Q = 1450 (dashed) to a Gaussian input pulse (solid).
  The output pulse is the input pulse convoluted with the impulse response
  of the oscillator. The symbols represent the resulting decay of the intensity in the cavity, as obtained from
  the FDTD calculation. } }\label{081024_GResponseQ1500Pulse0p2}
  \end{center}
\end{figure}

 The response of a damped harmonic oscillator with  \emph{Q} = 1450 to a Gaussian input
pulse with a width of $\tau_{ip}=0.2$ ps is shown in figure
\ref{081024_GResponseQ1500Pulse0p2}, together with the Gaussian
input pulse and the cavity response as calculated with FDTD. No
dispersion and no absorption was taken into account for the
FDTD calculations. In the harmonic oscillator case and in the
FDTD case the intensity decays exponentially and with the same
rate. Therefore, we conclude that the harmonic oscillator is a
suitable model to describe in a simple way the decay of the
microcavity. Furthermore the quality factor of the cavity
without absorption and dispersion is equal to \emph{Q} =
$1450\pm100$.

\begin{figure}[htb]
  \begin{center}
  \includegraphics[width=8cm]{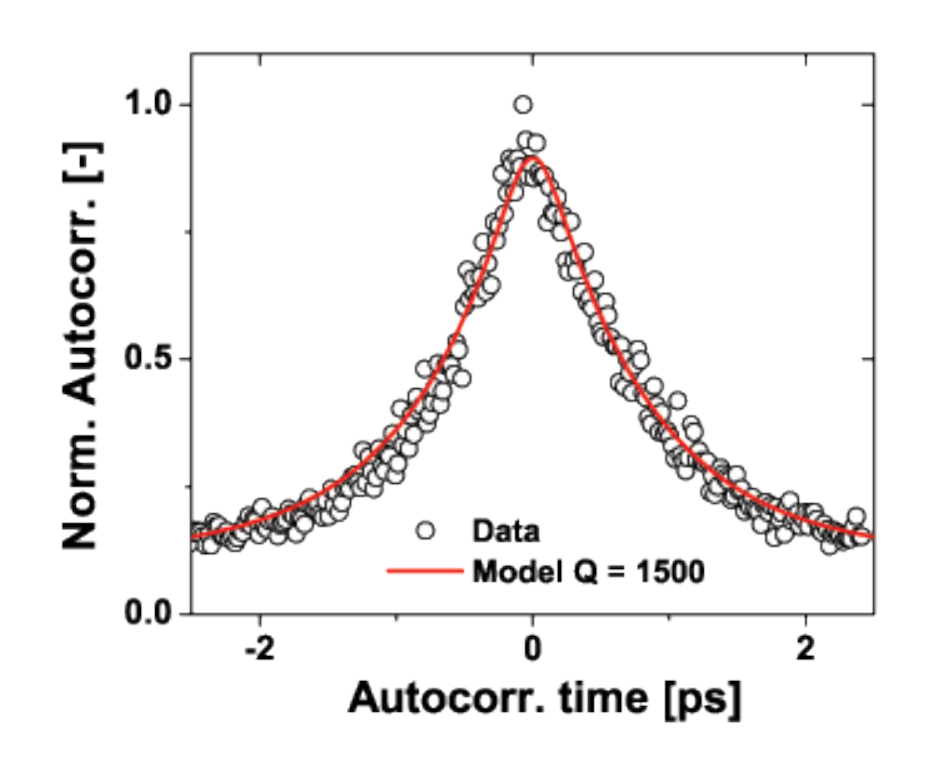}
  \caption{\emph{Autocorrelate of pulse transmitted at the cavity resonance (circles). The solid
  line represents the result from a simple oscillator model with Q = 1500. The input pulse duration is 0.12 ps.
  }}\label{081017_ACF_Cavity_Inc_Fit}
  \end{center}
\end{figure}

With the damped harmonic oscillator model we have calculated
the autocorrelates that are shown together with the measured
data in figure \ref{081017_ACF_Cavity_Inc_Fit} for a quality
factor of the damped harmonic oscillator \emph{Q} = 1500 and
Gaussian input pulse with width 0.12 ps.
Figure \ref{081017_ACF_Cavity_Inc_Fit} shows a very good
agreement between the measured autocorrelation trace and the
calculated autocorrelation trace for \emph{Q} = 1500. The
autocorrelate obtained from this model has a FWHM of 1.2 ps,
which is in very good agreement with the measured value of 1.1
ps.

Table \ref{tbl:invlinewidths} presents the results of the
various methods to determine the quality factor presented in
this work. We see that the value for the inverse linewidth
found from the transfer matrix model agrees well with the value
found from the autocorrelator measurement. The slight
discrepancy might be attributed to minor irregularities in the
structure. We conclude that the quality factor of the cavity is
$1500\pm100$ and the storage time of light $0.78\pm0.05$ ps.
The inverse linewidth measured in reflection and transmission
with white light spectroscopy is much smaller than the value
from autocorrelator measurements. The difference results from
inhomogeneous broadening due to the spatial gradient in the
thickness of the cavity layer. For a quality factor of 1500 we
expect a width of 0.64 nm, while we measure a width of
$1.2\pm0.1$ nm. Because of the focus diameter of 100 $\mu m$
and the spatial gradient of 5.64 nm/mm, we expect a broadening
of 0.56 nm. We find a total width of $1.2\pm0.1$ nm if we add
the broadening to the unbroadened width. The total width of
$1.2\pm0.1$ nm is in perfect agreement with the measured value
of $1.2\pm0.1$ nm. The FDTD calculation agrees very well with
the measured value and the value obtained from the transfer
matrix calculation. No absorption is taken into account in the
FDTD calculation, which is the case for the TM model.
\begin{table}
\begin{center}
\caption{Overview of inverse linewidths and cavity lifetimes
measured and calculated with presented methods.}
\begin{tabular}{@{}l|l|l|l@{}}
\multicolumn{4}{c}{}
\\\cmidrule(r){1-4}
 &Method & $\frac{\lambda_0}{\Delta\lambda} $ & $\tau_{cav}$ (ps)\\ \midrule[0.5pt]
Measurement & Autocorrelator & 1500 $\pm100$&  0.78$\pm0.05$\\
& Refl./Trans. & 830 $\pm50$& 0.43 $\pm0.02$ \\\midrule[0.5pt]
Theory& Transfer Matrix & 1640 $\pm100$& 0.93$\pm0.05$ \\
&FDTD & 1450 $\pm100$& 0.75 $\pm0.05$ \\
\end{tabular}
\label{tbl:invlinewidths}
\end{center}
\end{table}


The inverse relative linewidth of 830 that we find from the
transmission and reflectivity measurements is the result of
inhomogeneous broadening of the resonance. The origin of this
inhomogeneous broadening is most likely the spatial gradient in
the $\lambda$-thick GaAs layer. Because there is inhomogeneous
broadening, the planar microcavity should be viewed as a static
ensemble of microcavities of which each resonance frequency is
slightly shifted \cite{Lagendijk1993aa}. This is in agreement
with the results presented in \cite{de_Martini1990aa}, where
the spatial extent of modes was investigated. The effective
radius of the mode is given by $r_{eff}^2 =
(Q/2\pi)(\lambda/n)^2$. In our case we find a spatial extent in
the order of 5 $\mu m$, which is much smaller than the diameter
of the probe beam. In general our results show that the true
quality factor of a planar microcavity indeed can only be
obtained from a time-resolved measurement.

\section{Conclusion}

In the case of an inhomogeneously broadened resonance we have
shown that the intensity autocorrelate can be used to determine
the storage time of a cavity resonance. For an inhomogeneously
broadened microcavity resonance we have measured both the
spectral width and the intensity autocorrelation trace. The
intensity autocorrelation trace yields a value of the quality
factor that agrees well with the values found from transfer
matrix and FDTD calculations. The spectral width is affected by
inhomogeneous broadening and leads to the wrong value for the
quality factor.


\section*{Acknowledgments} \label{sec:Acknowledgement}
This research was supported by NanoNed, a nanotechnology
programme of the Dutch Ministry of Economic Affairs, and by a
VICI fellowship from the "Nederlandse Organisatie voor
Wetenschappelijk Onderzoek" (NWO) to WLV. This work is also
part of the research programme of the "Stichting voor
Fundamenteel Onderzoek der Materie" (FOM), which is financially
supported by the NWO.

\end{document}